\documentstyle[12pt]{article}

\newcommand{\BEQ}{\begin{equation}}    % Gleichungen Anfang ..
\newcommand{\BEA}{\begin{eqnarray}}
\newcommand{\EEQ}{\end{equation}}      % .. und Ende
\newcommand{\EEA}{\end{eqnarray}}
\newcommand{\rar}{\rightarrow}                   % Pfeil nach rechts
\newcommand{\ket}[1]{\left|#1\right\rangle}      % Ket-Zustand
\newcommand{\vm}[1]{\check{#1}}                  % Hut verkehrt fuer Math.
\newcommand{\zeile}[1]{\vskip #1 \baselineskip}  % N Zeilen ueberschlagen
\newcommand{\vekz}[2]
     {\mbox{${\begin{array}{c} #1  \\ #2 \end{array}}$}}
\newcommand{\build}[3]{\mathrel{\mathop{\kern 0pt#1}\limits_{#2}^{#3} }}
\catcode`\@=11
\def\numberbysection{\@addtoreset{equation}{section}
        \def\theequation{\arabic{equation}}}
\numberbysection

\begin{document}
%
%Titelseite
%
\begin{titlepage}
\null
hep-th/9306116
\vskip 0.5cm
\begin{center}
%~\hfill December 1992\\
%\vskip 0.5in
{\Large \bf Reaction-Diffusion Processes as Physical Realizations of
Hecke Algebras}
\vskip 0.5in
Francisco C. Alcaraz \footnote{Permanent adress:
Departamento de F\'{\i}sica,
Universidade Federal de S\~{a}o Carlos, 13560 S\~{a}o Carlos SP, Brasil}
and Vladimir Rittenberg
 \\[.3in]
{\em Physikalisches Institut, Universit\"at Bonn \\
Nu{\ss}allee 12, D - 5300 Bonn 1, Germany}
\zeile{1}
\end{center}
%
% Abstract
%
\begin{abstract}
%{\bf Abstract} \zeile{1}
We show that the master equation governing the dynamics of simple diffusion
 and certain chemical reaction processes in one dimension give time
evolution operators (Hamiltonians) which are realizations of Hecke algebras.
In the case of simple diffusion one obtains, after similarity transformations,
reducible hermitian representations while in the other cases they are
non-hermitian and correspond to supersymmetric quotients of Hecke
algebras.

\end{abstract}
%\zeile{3}

\end{titlepage}

%\newpage
%
%Text der Arbeit
%

\section*{}
%%%%%%%%
        It is well known in the literature that several integrable quantum
chains corresponding to magnetic systems \cite{R1} can be represented as
generators of  Hecke algebras $H_n(q)$, when certain artificial
interactions are added in the bulk and the surface.
In this letter we show that the master equation describing the
dynamics of some chemical processes limited by diffusion give
representations of Hecke algebras, where the supplementary interactions
appear naturally.

The Hecke algebra $H_n(q)$ (with $n = L-1$) is an associative algebra
with generators $e_i$
($i=1,\ldots,L-1$) satisfying the relations
\BEA
 e_i e_{i\pm 1} e_i - e_i = e_{i\pm 1} e_i e_{i\pm 1} - e_{i\pm 1}
\label{1}\\
 \left[ e_i , e_j \right] =0 \;\; ; \;\; |i-j| \geq 2 \label{2} \\
 e_{i}^{2} = \left( q + q^{-1} \right) e_{i}  \label{3}
\EEA
where $q$ is in general a complex parameter. In our applications $q$ is
always  real. The main interest in this algebra in
statistical mechanics is due to the fact that a spectral-dependent
 $\vm{R}(u)$ matrix, satisfying the Yang-Baxter relations
\cite{R2}, can be constructed in a standard form. This construction,
also called
``Baxterization'' \cite{R3}, implies that the Hamiltonian
$\sum_{i=1}^{L-1} e_i$  has an
infinite number of conserved charges and this is the first step towards
the exact integrability of $H$.

One can define \cite{R4} a sequence $(P,M)$ of quotients algebras of
$H_q(n)$ ($P$ and $M$ are positive integers). Since the difinition is quite
complicated, we will  give only a few examples. The quotient $(1,0)$
implies taking $e_1 = 0$. This quotient is not relevant for our purposes.
The $(2,0)$ quotient is the Temperley-Lieb algebra:
\BEQ \label{4}
e_{i} e_{i\pm 1}e_{i} = e_{i}.
\EEQ
The $(1,1)$ quotient reads:
\BEQ \label{4p}
\left( e_{i}e_{i+2}\right) e_{i+1} \left(q+q^{-1}-e_{i}\right)
\left(q+q^{-1}-e_{i+2}\right) =0 .
\EEQ
For our purposes it is sufficient to mention that each $(P,M)$ quotient
has as a representative one of the $(P,M)$ Perk-Schultz \cite{R5} quantum
chains where the Hamiltonian is given by the sum of $H_n(q)$ generators:
\BEQ \label{19p}
H^{(P,M)} = \sum_{j=1}^{L-1} e_j^{(P,M)}
\EEQ
where
\BEA
e_j^{(P,M)} &=& \frac{q+q^{-1}}{2} -
\left( \sum_{\alpha \neq \beta}
 E_j^{\alpha \beta} E_{j+1}^{\beta \alpha} + \frac{q+q^{-1}}{2}\sum_{\alpha
=0}^{N-1} \epsilon_{\alpha}E_j^{\alpha \alpha}E_{j+1}^{\alpha \alpha}  \right.
\nonumber \\
&& \left.
+ \frac{q-q^{-1}}{2}\sum_{\alpha \neq \beta} \mbox{\rm sign}(\alpha-\beta)
E_j^{\alpha \alpha}  E_{j+1}^{\beta \beta} \right). \label{19}
\EEA
with $N = P + M$. The $N \times N$ matrices $E^{\alpha \beta}$ have elements:
\BEQ \label{8p}
(E^{\alpha \beta})_{\gamma \delta} = \delta_{\alpha \gamma} \delta_
{\beta \delta} \quad (\alpha, \beta = 0, 1, \ldots, N-1)
\EEQ

and
\BEQ \label{9p}
\epsilon_o = \epsilon
_1 = \ldots = \epsilon_{P-1} = - \epsilon_P = \ldots = -
\epsilon_{P+M-1}.
\EEQ
The quantum chains $H^{(P,M)}$ are $U_q(SU(P/M))$ invariant \cite{R6} and
have the important property that they contain all the irreducible
representations of the $(P,M)$ quotient. In the examples we are going to
mention, if a chain corresponds to a certain quotient $(P,M)$ them its energy
levels (not the degeneracies!) are the same as those of the representative
of this quotient given by Eqs.(\ref{19p}-\ref{19}). We now turn to the
physical problem.

It was pointed out recently \cite{R7}, that the one-dimensional master
equation corresponding to a chain with two-body transition rates can be
written as a Schr\"odinger equation with a Hamiltonian described by a
quantum chain with nearest-neighbour interactions. Several examples
were found in which this Hamiltonian had precisely the structure of
Eqs.(\ref{19p}-\ref{19}). In this letter we present this problem in
detail. We will use the notations of Ref.\cite{R7}.

        In order to describe the dynamics of general reaction-diffusion
\cite{R7}
processes we use a master equation.
Let us consider an open chain with $L$ sites, where at each site
$i =1,2,\ldots,L$
we attach a variable
 $\beta$ taking $N$ integer
values ($\beta = 0,1,\ldots,N-1$).
By convention we attach the value $\beta =0$
to a vacancy (inert state) and $\beta = 1,\ldots,N-1$ a molecule of
type $\beta$. The master equation describing the time evolution of
 the probability distribution
$P(\{\beta\};t)$ is
\BEA
\lefteqn{ \frac{\partial P(\{\beta \};t)}{\partial t} =
\sum_{k=1}^{L-1} \left[ - w_{0,0} (\beta_k ,\beta_{k+1})
P(\beta_1,\ldots ,\beta_L ;t) \vekz{ }{~} \right. } \nonumber \\
& & + \left. {\mathop{{\sum}'}_{\ell,m=0}^{N-1}}
w_{\ell,m}(\beta_k,\beta_{k+1}) P(\beta_1,\ldots,[\beta_k +\ell]_N,
[\beta_{k+1}+m]_N,\ldots,\beta_L ;t) \right] \label{5}
\EEA
where $w_{\ell,m}$ are the transition
rates and the prime in the second sum
indicates the exclusion of the pair $\ell =m=0$. In (\ref{5})
 the symbol
$[x + y]_N$ means the addition $(x + y)$,modulo $N$. We restrict
ourselves to the
cases where the transition rates are homogeneous and depend only on the
molecules on nearest neighbouring sites. The transition rate $w_{\ell,m}
(\alpha,\beta)$ gives the probability, per unit of time, that an initial
state where two consecutive sites are occupied by molecules of type $(
\alpha+\ell,\beta+m)$ end up occupied by molecules of type $(\alpha,\beta)$.
The rate $w_{0,0}(\alpha,\beta)$ gives the probability, per unit of time, that
the configuration of two neighbouring sites $(\alpha,\beta)$ are changed. From
these definitions the conservation of probabilities implies
\BEQ \label{6}
w_{0,0}(\alpha,\beta) = {\sum_{r,s}}' w_{r,s}(\alpha-r,\beta-s),
\EEQ
where $r=s=0$ is again excluded.

The master equation (\ref{5}) can be interpreted as a Schr\"odinger
equation
\BEQ \label{7}
\frac{\partial}{\partial t} \ket{\Psi}= - {H} \ket{\Psi}
\EEQ
in Euclidean time, if we identify the probability $P(\{\beta\},t)$ as the
wave function.  The Hamiltonian in (\ref{7})
\BEQ \label{8a}
H = \sum_{j=1}^{L-1} H_j
\EEQ
acts in a Hilbert space of dimension $N^L$ , while $H_j$ acts in the subspace
 $V^{(j)} \otimes V^{(j+1)}$  and is given by
\BEQ \label{8b}
H_j = U_j - T_j,
\EEQ
where
\BEA
T_j &=&{\mathop{{\sum}'}_{\ell,m=0}^{N-1}} \sum_{\alpha,\beta=0}^{N-1}
w_{\ell,m}(\alpha,\beta)  E^{\alpha,[\alpha+\ell]_N}  \otimes
 E^{\beta,[\beta+m]_N} \label{8c} \\
U_j &=&\sum_{\alpha,\beta=0}^{N-1} w_{0,0}(\alpha,\beta)
E^{\alpha \alpha} \otimes E^{\beta \beta} \label{8d}.
\EEA
The above
Hamiltonian  describes the time evolution from an initial probability
$\Psi_0(\{\beta\},t=0)$ with dynamical processes given by the transition
rates $w_{\ell,m}(\alpha,\beta)$.

Here are some examples of processes in which the $H_j$'s of Eq.(\ref{8a})
coincide with the generators $e_j$ of the Hecke algebra $H_n(q)$.

a) {\bf diffusion processes:} we consider $(N-1)$ types of molecules
$(A_1,\ldots,A_{N-1})$ which diffuse to the right:
\BEQ \label{17}
A_b + \emptyset \rar \emptyset + A_b \quad (b =1, \ldots,N-1)
\EEQ
with equal rates $\Gamma_R$ and to the left:
\BEQ \label{18}
\emptyset + A_b \rar A_b + \emptyset \quad (b =1, \ldots,N-1)
\EEQ
with equal rates $\Gamma_L$. Choosing the time scale in such a way
that the diffusion
constant $D=\sqrt{\Gamma_L \Gamma_R} = 1$ and introducing
 $q=\sqrt{\Gamma_R /
\Gamma_L}$, up to an equivalence transformation, we obtain
 $H_j = e_j$ corresponding
to the quotient $(2,0)$.

b) {\bf diffusion and interchange processes:} to the diffusion processes
(\ref{17}) and (\ref{18}) we add the interchange to the right processes:
\BEQ \label{19l}
A_b + A_c \rar A_c + A_b \quad  (b \neq c, b > c)
\EEQ
with rates $\Gamma_R$ and with rates $\Gamma_l$ if $b < c$. In this case
the quotient is $(N,0)$ and the obtained representation of $H_n(q)$
is hermitian (the same applies to the diffusion processes of case a).

c) {\bf diffusion, interchange  and coagulation processes:} we now add to the
processes (\ref{17}-\ref{19l}) only one reaction for the molecule $A_1$
(the molecules are ordered!)
\BEQ \label{20}
A_1 + A_1 \rar \emptyset  + A_1
\EEQ
with a rate $\Gamma_R$ and the processes where $\emptyset$ and $A_1$
interchange their position in the final state with rate $\Gamma_L$. In this
case the quotient is $(N-1,1)$ and the representation of $H_n$ is
non-hermitian. As a rule, non-hermitian representations occur for
quotients $(P,M)$ with $M \neq 0$ i.e. when the representative Hamiltonians
$H^{(P,M)}$ are supersymmetric.

One obtains new quotients of $H_n$ generalizing the reaction (\ref{20}) by
taking
\BEQ \label{21a}
A_r + A_r \rar A_s + A_r \quad  (\mbox{rate} \; \Gamma_R)
\EEQ
\BEQ \label{21b}
A_{r'} + A_{r'} \rar A_{s'} + A_{r'} \quad (\mbox{rate} \; \Gamma_R)
\EEQ
where $s = r \pm 1$, $s' = r' \pm 1$, and $r \neq r',s';s \neq r',s'$. In
this way one obtains the quotient $(N-2,2)$. Taking more reactions like
in Eq.(\ref{21a},\ref{21b}) one can get arbitrary quotients $(P,M)$.

d) {\bf diffusion-polymerisation (modulo N) processes:} in this case
besides the diffusion (Eqs. \ref{17}-\ref{18}), one considers the processes:
\BEQ \label{22a}
A_r + A_s \rar \emptyset + A_{[r +s]_N} \quad (\mbox{rate} \; \Gamma_R, \;
[r + s]_N \neq 0)
\EEQ
\BEQ \label{22b}
A_r + A_s \rar A_{[r+s]_N} +  \emptyset  \quad (\mbox{rate} \; \Gamma_L, \;
[r + s]_N \neq 0)
\EEQ
\BEQ \label{22c}
A_r + A_s \rar \emptyset + \emptyset \quad (\mbox{rate} \;
\Gamma_R + \Gamma_L,\; [r + s]_N = 0) \quad
(r,s = 1,2, \ldots, N-1).
\EEQ
For all these processes one obtains the quotients $(1,1)$. Let us remark
that the processes described by Eq. (\ref{22a}-\ref{22c}) were already
considered in
Ref.\cite{R9} and called "P-poly" for irreversible diffusion cluster-cluster
aggregation processes. Using probabilistic methods several exact
results were obtained (see also Ref.\cite{R10}) concerning the concentrations
of the different types of molecules. We believe that the profound reason
which explains why these results could have been obtained is that the $(1,1)$
quotient has a spectrum which can be obtained using free fermion methods
\cite{R11}) (see also \cite{R12}).

e) {\bf diffusion, interchange and annihilation processes:} because of
interchange processes, the labels of the molecules are ordered; we can
now add the following two processes:
\BEQ \label{23}
A_m + A_m \rar A_{m+1} + A_{m+1} \quad   (\mbox{rate} \; \Lambda_1)
\EEQ
\BEQ \label{24}
A_m + A_m \rar A_{m-1} + A_{m-1} \quad   (\mbox{rate} \; \Lambda_{-1}),
\EEQ
with $\Lambda_1 + \Lambda_{-1} = \Gamma_R + \Gamma_L$. In this way we obtain
again the quotient $(2,1)$.

The reader might wonder how the above $H_n(q)$ quotients were obtained. From
the work done in Ref.\cite{R7} we have developped a certain intuition for
chemical processes up to $N = 3$. We have then used the computer to check
that no other solutions exist for $N = 2$ and $3$ and generalized
the solutions obtained for $N$ up to $3$ to all $N$ and checked them. There
are certainly other solutions which one can discover studying systems with
$N = 4$ and certainly  examples of the quotients $(2,2)$ different from the
 the one given in Eq.(\ref{21a},\ref{21b}) will show up. We hope
that the present effort
to connect representatives of $H_n(q)$ and chemical processes will help
to compute
various correlation functions for the latter.
One immediate consequence of the above mentioned connecxion is that all
processes with $q \neq 1$ are massive (the concentrations for example have
an exponential fall-off in time) and all those with $q =1$ are massless
(algebraic fall-off).

\zeile{3}
\noindent{\bf Acknowledgements}
\zeile{2}
F.A. thanks the Conselho Nacional de Desenvolvimento Cient\'ifico and
Tecnol\'ogico - Brasil
and the Deutsche Forschungsgemeinschaft - DFG - Germany for support.

%\newpage

\end{document}